\documentclass[aps,twocolumn,prl,showpacs,amssymb]{revtex4}

\usepackage{graphicx}
\usepackage{dcolumn}
\usepackage{bm}

\begin{document}
\title{On optical black holes in moving dielectrics}
\author{V. A. De Lorenci}
 \email{delorenci@unifei.edu.br}
\author{R. Klippert}
 \email{klippert@unifei.edu.br}
\affiliation{Instituto de Ci\^encias,
Universidade Federal de Itajub\'a, 
Av.\ BPS 1303 Pinheirinho, 37500-903 Itajub\'a, MG, Brazil} 
\author{Yu.\ N. Obukhov}
 \altaffiliation[On leave of absence from the ]{Moscow 
                           State University}
 \email{yo@ift.unesp.br}
\affiliation{Instituto de F\'\i sica Te\'orica, 
Universidade Estadual Paulista, Rua Pamplona 145 Paulista, 
01405-900 S\~ao Paulo, SP, Brazil}
\date{\today}

\begin{abstract}
We study the optical paths of the light rays propagating inside 
a nonlinear moving dielectric media. For the rapidly moving dielectrics 
we show the existence of a distinguished surface which resembles, as far 
as the light propagation is concerned, the event horizon of a black hole.
Our analysis clarifies the physical conditions under which electromagnetic 
analogues of the gravitational black holes can eventually be obtained in 
laboratory.  
\end{abstract}
\maketitle


One of the most striking and well established results of 
general relativity theory is the prediction of a bounded region in 
space from where nothing can escape, not even light. Such an 
object, a black hole, arises as the spherically symmetric
solution of Einstein equations, and its analysis on the basis
of the semi-classical quantum field theory was considered to
be fundamental for the understanding of quantum gravity.
In particular, in the context of the semi-classical gravity, 
Hawking suggested that black holes can evaporate \cite{Hawking75}. 
Since the temperature of the Hawking radiation is very low, 
astrophysical events are unlikely to give an evidence for 
such a phenomenon. 

During the last decades efforts have been put into a development 
of the alternative physical models in which such a prediction could 
be experimentally verified under laboratory conditions. This class 
of models was called `analog gravity' \cite{conference}, and it 
deals with the propagation of sound waves in moving fluids 
\cite{Unruh,Unruh95,Visser98,Garay} 
and with the electrodynamics of nonlinear media 
\cite{comm,Brevik,analog,Souza,Salim,PRL}, among others 
(see for instance Ref.\ \onlinecite{Barcelo} and references therein).

The search for electromagnetic black holes represents an
interesting topic in the broad area of analogue gravity, 
but the effective construction of such structures still
remains an open problem. In a recent work \cite{Obukhov} 
a certain progress has been achieved in this direction.
In particular, it was shown that the birefringence effect 
occurring in moving nonlinear dielectrics 
can be described in terms of the optical geometries 
(also known as the electromagnetic effective geometries). 
Accordingly, the propagation of the electromagnetic light rays 
is governed by the optical metric tensors:
\begin{eqnarray}
g^{\mu\nu}_-=&&\!\!\!\eta^{\mu\nu}+(\epsilon\mu-1)u^\mu u^\nu
-\mu u^{(\mu}e^{\nu)\alpha}F_{\alpha\beta}u^\beta
\nonumber\\&&\!\!\!
-\,\frac{1}{\epsilon}\,u_\alpha e^{\alpha(\mu}F^{\nu)\beta}u_\beta,
\label{g-}\\
g^{\mu\nu}_+=&&\!\!\!\eta^{\mu\nu}+(\epsilon\mu-1)u^\mu u^\nu.
\label{g+}
\end{eqnarray}
Here $\eta^{\mu\nu}$ is the background (flat) spacetime metric, 
and the magnetic permeability $\mu$ is assumed to be a constant.  
The derivative $e^{\mu\nu}\doteq\partial\epsilon/\partial F_{\mu\nu}$ 
of the dielectric permittivity $\epsilon$ with respect to the 
electromagnetic field measures the nonlinear effects. The 
motion of the medium is described by the velocity vector $u^\mu$ 
measured in units of the velocity of light in vacuum 
({\em i.e.}, $c=1$ except specified otherwise). 
The meaning of the Eqs.\ (\ref{g-}),(\ref{g+}) 
is that the wave vectors $K_\lambda=\partial_\lambda\Sigma$ 
lie on the null geodesics of the above two optical metrics, 
depending on their polarization state.  
Here $\Sigma$ denotes the corresponding wave-front surface.

The aim of the present work is to show explicitly how the analogue 
event horizons can be built from the above metric structures.  
We analyze the propagation of the two light rays 
described by the optical metrics (\ref{g-}),(\ref{g+}).  
Despite of birefringence, the horizons 
for both polarization modes are shown to coincide.  
The spherical configuration is discussed, and we find an analogue 
event horizon resembling the Schwarzschild event horizon. 
Further, we study the cylindrically symmetric configuration as 
a possibly simpler structure to be built in a laboratory.  


Let us now examine the light propagation inside a moving nonlinear 
dielectric matter. Consider at first the spherically symmetric case
with the background metric 
$ds^2_\eta=dt^2-dr^2-r^2d\theta^2-r^2\sin^2\theta\,d\varphi^2$ 
in which the dielectric flows radially with the velocity 
$u^\mu=\gamma\,(1,\beta,0,0)$, where $\gamma\doteq (1-\beta^2)^{-1/2}$.
We assume a point charge fixed at the origin, and correspondingly
the only non-zero component of the electromagnetic field is then 
$F^{01}=E_r$. Thus $e^{01}F_{01}=E(d\epsilon/dE)\doteq\epsilon'$, 
with $E=|E_r|$.  The optical metric (\ref{g-}) then reads 
\begin{eqnarray}
g^{00}_- &=& (\epsilon + \epsilon')\,\gamma^2
\left(\mu - \frac{\beta^2}{\epsilon}\right),\label{g00}\\
g^{01}_- &=& (\epsilon + \epsilon')\,\gamma^2\beta
\left(\mu - \frac{1}{\epsilon}\right),\label{g01}\\
g^{11}_- &=& (\epsilon + \epsilon')\,\gamma^2\left(
\beta^2\mu - \frac{1}{\epsilon}\right),\label{g11}\\
g^{22}_- &=& \sin^2\theta \, g^{33}_- \, = \,-\,\frac{1}{r^2} . 
\label{g33}
\end{eqnarray}
Note that $g^{\mu\nu}_+=g^{\mu\nu}_-\vert_{\epsilon'=0}$.  
For spherically symmetric flows $\beta=\beta(r)$, the two metrics 
$g^{\mu\nu}_\pm$ evidently admit the two Killing vector fields 
$\xi=\{\partial/\partial t,\partial/\partial\varphi\}$.  
Written in components, $\xi^\mu_a=\delta^\mu_a$ with $a=0,3$.  
The light then propagates along the null geodesics with the 
tangent vectors $K^\mu=\dot x^\mu=(\dot t,\dot r,\dot\theta,
\dot\varphi)$ in such a way that $\xi_0^\mu K_\mu = K_0$ and 
$\xi_3^\mu K_\mu = K_3$ are constants along the corresponding 
optical path. Here the `dot' denotes the derivative with
respect to the affine parameters $s$ along the optically null 
geodesic lines. For simplicity, we choose the radial propagation 
with $K_2=0=K_3$. The path of the light ray is then determined by 
$\dot{t}=\dot{r}(n\pm\beta)/(n\beta\pm1)$.  
Here $n = \sqrt{\mu\epsilon}$ is the refraction index of the medium 
in the absence of external fields.
The plus (minus) sign corresponds to the wave propagating from (to) 
the origin. 
The radial coordinate of the photon then changes in time as 
\begin{equation}
\frac{dr}{dt}= {\frac {n\beta\pm 1}{n\pm \beta}}
\label{drdt}.
\end{equation}
For the case of the matter flow directed {\it outwards}, 
i.e., $\beta>0$, 
then the {\it incoming} light (with $\dot{r}<0$) apparently 
``freezes'' at the radius $r_h$ such that $n\beta\vert_{r_h}=1$. The
same happens for an {\it outgoing} light in the {\it inwardly} directed
flow: the ``freezing'' occurs then at $n\beta\vert_{r_h}= - 1$. Both cases
are compactly summarized in 
\begin{equation}\label{hor}
n^2\beta^2\,\vline\,{\hbox{\raisebox{-1.5ex}{\scriptsize{$r_h$}}}} 
= 1.
\end{equation}
It is worthwhile to note that ``freezing'' is absent for the light 
propagating in the {\it same} direction 
as the moving dielectric medium.

There is a clear physical interpretation of the above result. Recall that
$\beta = v/c$ for the 3-velocity $v$ of the matter flow, and $v_c = c/n$ 
is the velocity of light in a medium with refraction index $n$ in the 
absence of external fields. Then
we can straightforwardly recast the equation (\ref{drdt}) into the form 
of the relativistic transformation of the velocity:
\begin{equation}
\frac{dr}{dt}= {\frac {v\pm v_c}{1\pm vv_c/c^2}}.\label{drdt1}
\end{equation}
The `freezing' of light takes place at a surface on which the velocity 
of matter $v$ becomes equal to the velocity of light $v_c$ in the medium.  
The geometrical interpretation is also clear: we find that the 
metric coefficient $g^{11}_-(r_h)=0$ vanishes at the radius $r_h$, 
see Eq.\ (\ref{g11}). The same conclusion is true also for the second 
optical metric, $g^{11}_+(r_h)=0$, i.e., both polarization modes 
(\ref{g-}) and (\ref{g+}) behave similarly. For the radial motion 
considered above, the stationary metrics given in the Eqs.\ (\ref{g-}), 
(\ref{g+}) can be rewritten into the static ``black hole'' form 
\begin{equation}
ds_{\pm}^2 = g_{00}^\pm\,d\tilde{t}^2 - {\frac {d\tilde{r}^2} {g_{00}^\pm}}
- r^2d\theta^2 - r^2\sin^2\theta\,d\varphi^2,
\end{equation}
with $g_{00}^-={\gamma^2(1-n^2\beta^2)}/{\mu(\epsilon+\epsilon')}$, 
by means of the coordinate transformations 
\begin{eqnarray}
d\tilde{t} &=& dt + \frac{g^{01}_\pm}{g^{11}_\pm}\,dr, \\
d\tilde{r} &=& dr\,\sqrt{(g^{01}_\pm)^2-g^{00}_\pm g^{11}_\pm}. 
\end{eqnarray}
We note, for the case here described, 
that $g_{00}^\pm$ is proportional to $g^{11}_\pm$.

{}From the above analysis, which generalizes the previous calculations 
\cite{PRL}, we find that either an ultra-relativistic motion of the 
matter or a highly refringent medium \cite{Kocharovskaya,Phillips}, 
as it is found in Bose-Einstein condensates (see, however, a relevant 
discussion in Ref.\ \onlinecite{Brevik}),
is needed in order to display such a horizon structure.  
Highly refringent media are usually quite dispersive, and thus the 
effective horizon would occur in this case only for a narrow 
range of frequencies.  A similar feature occurs also for sonic 
analog models, where the horizon is defined only in a narrow 
range of low frequencies.  
A practical realization of the analogue black hole configuration 
requires that either $n=n(r)$ or $\beta=\beta(r)$, or even both.  

Recently, it has been proposed that Hawking temperature 
is a purely kinematic effect that is generic to Lorentzian
geometries containing event horizons \cite{Visser98b}, 
and thus being dependent only on the effective metric structure.  
For the spherically symmetric solution the Hawking temperature $T$ 
of the analogue black hole is straightforwardly calculated 
for the metric (\ref{g00})-(\ref{g33}) as
\begin{equation}
T = \left|{\frac {\hbar\gamma^2\beta n}{2\pi k_{\scriptscriptstyle B}}}
\left(\frac{\partial\beta}{\partial r}
+\frac{\beta\epsilon'}{2\epsilon E}
\frac{\partial E}{\partial r}\right)\right|_{r_h}
\,.\label{temp} 
\end{equation}
Equation (\ref{temp}) shows that the nonlinearity 
may significantly contribute to this temperature.  
It should be noted, however, that the complete understanding of 
the physical meaning of the Hawking temperature associated with an
analogue black hole is still an open question \cite{Brevik}, and it can 
be settled only after a detailed analysis of the radiation 
processes at the horizon.  It should be stressed that 
the approximation of geometrical optics becomes unreliable 
for the discussion of modes whose wavelengths are comparable 
with the size of the horizon, as occurs in Hawking radiation processes.

Experimentally though, it appears to be a rather difficult task to 
maintain a stationary spherically symmetric and inhomogeneous flow.  
In order to exhibit a more realistic configuration, we will now focus 
on the cylindrical symmetry. A particular case of such a configuration
is the vortex matter flow which was discussed recently 
\cite{leo,marklund}. Let us consider a more general situation of a 
rotating dielectric body subjected to an electric field $E_z$ 
directed along the axis $z$ of rotation 
\footnote{Such electromagnetic field configuration 
corresponds to the experimental situation of a rotating condenser.}.  
The background metric is then of 
the form $ds^2_\eta = dt^2-d\rho^2-\rho^2d\varphi^2-dz^2$. The 
electromagnetic field has the only non-zero component $F^{03}=E_z$ 
and we denote $e^{03}F_{03}=E(d\epsilon/dE)\doteq\epsilon'$, 
where $E=|E_z|$.  
For the matter 4-velocity we have $u^\mu=\gamma(1,\beta,\omega,0)$, 
with $\gamma=1/\sqrt{1 -\beta^2-\rho^2\omega^2}$, 
while $\omega$ and $\beta$ being arbitrary functions 
of the radial coordinate $r$.  
The optical geometry (\ref{g-}) now reads 
\begin{eqnarray}
g^{00}_-=&&
\gamma^2\left[\mu(\epsilon+\epsilon')-\beta^2-\rho^2\omega^2\right]
\label{g-00},\\
g^{02}_-=&&\frac{\omega}{\beta}g^{01}_-=
\gamma^2\omega\left[\mu\left(\epsilon+\frac{\epsilon'}{2}\right)
-1\right]
\label{g-02},\\
g^{11}_-=&&
\gamma^2\,[\mu\epsilon\beta^2+\rho^2\omega^2-1]
\label{g-11},\\
g^{12}_-=&& \gamma^2\beta\omega\,(\mu\epsilon-1)
\label{g-12},\\
g^{22}_-=&& \gamma^2\left[\mu\epsilon\omega^2 
- {\frac{1}{\rho^2}} + {\frac{\beta^2}{\rho^2}}\right]
\label{g-22},\\
g^{33}_-=&& -\,1-\frac{\epsilon'}{\epsilon}\gamma^2
\label{g-33}.
\end{eqnarray}
For the motion under consideration, 
the above metric components, Eqs.\ (\ref{g-00})--(\ref{g-33}), 
depend only on the radial coordinate $\rho$. Then, for this geometry we have 
the three Killing vectors $\partial/\partial t,\partial/\partial\varphi,
\partial/\partial z$. In components, $\xi^\mu_{a}=\delta^\mu_a$, with 
$a=0, 2, 3$. The null geodesic equations $K^\mu\nabla_\mu K_\lambda =0$ 
for the wave vector $K_\lambda$ in the optical metric $g^{\mu\nu}_-$ are 
then more easily solved by using the fact that $\xi^\mu K_\mu=
{\rm const}$ for each Killing vector $\xi^\mu$.  That is, 
the covariant components $K_0$, $K_2$, $K_3$ of the wave vectors 
$K^\mu=\dot x^\mu=(\dot t,\dot\rho,\dot\varphi,\dot z)$ 
are constant along the corresponding geodesic lines.  
The remaining component $K_1$ is obtained from the null 
condition $g^{\mu\nu}_-K_\mu K_\nu=0$, 
and the geodesic equation for the radial coordinate reads 
\begin{widetext}
\begin{equation}
\dot{\rho}=\pm\sqrt{(g^{10}_-K_0+g^{12}_-K_2)^2-g^{11}_-
[2g^{02}_-K_0K_2+g^{00}_-(K_0)^2+g^{22}_-(K_2)^2+g^{33}_-(K_3)^2]}
\label{dotrho},
\end{equation}
where the plus (minus) sign corresponds to 
a light ray propagating from (to) the symmetry axis $z$.  
Also, $\dot{z}=g^{33}_-K_3$ and 
\begin{eqnarray}
\dot{t}=&&\!\!\!\frac{[g^{00}_-g^{11}_--(g^{01}_-)^2]K_0
+[g^{02}_-g^{11}_--g^{01}_-g^{12}_-]K_2+g^{01}_-\dot{\rho}}
{g^{11}_-}
\label{tdot},\\
\dot{\varphi}=&&\!\!\!\frac{[g^{02}_-g^{11}_--g^{01}_-g^{12}_-]K_0
+[g^{11}_-g^{22}_--(g^{12}_-)^2]K_2+g^{12}_-\dot{\rho}}
{g^{11}_-}
\label{phidot}.
\end{eqnarray}
\begin{figure}[thp]
\leavevmode
\centering
\hspace*{-0cm}
\mbox{
\includegraphics[scale=0.93]{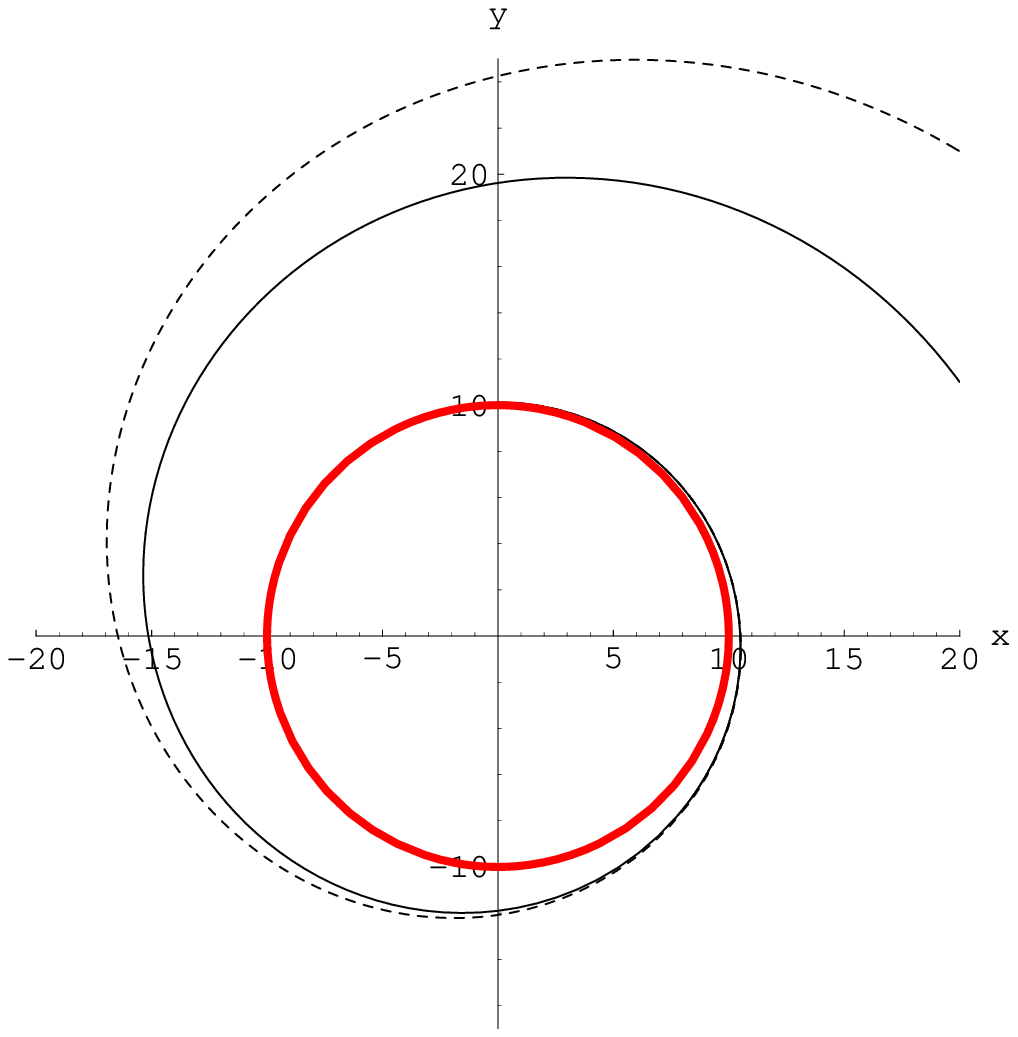}
\hspace{0.5cm}
\raisebox{9.8ex}{\includegraphics[scale=0.93]{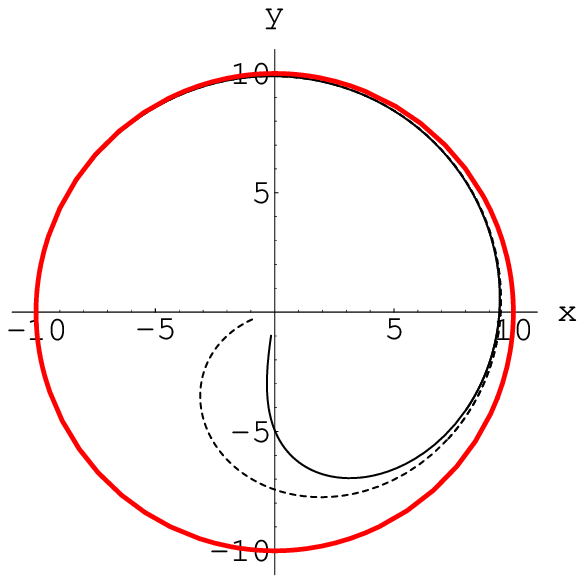}}
}
\caption{The light path for both polarization modes are shown
for the cylindrical (`black string') configuration.  
The left/right figure corresponds to a dielectric fluid 
flowing outwards/inwards, with radial velocity $\beta_{\rm out}(\rho)
=10^{-2}/\rho$ and $\beta_{\rm in}(\rho)=-10^{-4}\rho$, respectively.  
The solid lines depict the ordinary rays, whereas the dashed lines 
represents the extraordinary rays. 
For simplicity, we assume in the plots a rigid rotation 
with $\omega=10^{-4}$, and the refraction index is $n=10^3$.  
Besides, we take $\mu\epsilon'=10^{10}$ for the extraordinary rays.  
(The choice of a large value for $\epsilon'$ is but illustrative, 
since the typical values would produce the rays too close to be 
distinguished from one another in the plots.)  
The thicker lines correspond to the analogue horizons in both cases.  
\label{horizon}
}
\end{figure}
\end{widetext}
By specifying the dynamics of the matter, i.e. the functions $\beta$ 
and $\omega$, all the metric components $g^{\mu\nu}_-$ will be 
given functions of $\rho$. 
Provided $\rho=\rho(s)$ is obtained from Eq.\ (\ref{dotrho}), 
the three other coordinates are then found by quadrature.  

Let us now analyze the ordinary ray described by $g^{\mu\nu}_+$, 
which is obtained from the above computations when we set $\epsilon'=0$. 
For simplicity, we also consider the planar initial conditions for the light 
propagation such that $K_2=0=K_3$. Equations (\ref{dotrho}) and (\ref{tdot}) 
then yield 
\begin{equation}
\frac{d\rho}{dt}=
\frac{\gamma\left(n\beta+\sqrt{1-\rho^2\omega^2}\right)
\left(n\beta-\sqrt{1-\rho^2\omega^2}\right)}
{(n^2-1)\gamma\beta\mp\sqrt{n^2(1-\beta^2)-\rho^2\omega^2}}
\label{rho(t)}.
\end{equation}
As compared to the spherical case, Eq.\ (\ref{drdt}), 
the solution of Eq.\ (\ref{rho(t)}) 
displays the same qualitative behavior for the cylindrical configuration. 
The cylindrical horizon is located at $\rho=\rho_h$ defined from
\begin{equation}
\left(n\beta\pm\sqrt{1-\rho^2\omega^2}
\,\right)
\,\vline\,{\hbox{\raisebox{-1.5ex}{\scriptsize{$\rho_h$}}}}=0
\label{horizonrho}.
\end{equation}  
Purely vortical motion of the fluid (with $\beta=0$) 
does not produce an analogue event horizon \cite{comm}, 
since $\rho_h$ from Eq.\ (\ref{horizonrho}) 
would lie beyond the limit of applicability 
of a rigid body in special relativity.  

We note that $d\varphi/d\rho$ diverges at the surface defined
by Eq.\ (\ref{horizonrho}).  
As a result, the incoming/outgoing geodesic light rays spiral 
towards a horizon radius $\rho_h$, which can be explicitly 
demonstrated by the numerical integration of $d\varphi/d\rho$, 
see Fig.\ \ref{horizon}. Moreover, the condition of a horizon 
for the optical metric $g^{\mu\nu}_-$ is precisely given by the 
Eq.\ (\ref{horizonrho}).  In terms of the optical metrics, 
Eqs.\ (\ref{g-00})--(\ref{g-33}), the position
of the horizon is defined by the equation $g^{11}_\pm(\rho_h)=0$.  
In Fig.\ \ref{horizon}, we also depict the extraordinary ray which
presents a qualitative behavior quite similar to its ordinary counterpart.  
Although the calculations look somewhat more involved when the above 
initial conditions are not fulfilled, it can be shown that 
the horizon condition (\ref{horizonrho}) remains unchanged.  
Therefore, we conclude that, for a given stationary flow configuration, 
the analogue horizon structure is of geometrical nature, as soon as 
it depends neither on the initial conditions nor on the 
polarization of the propagating light rays.


Summarizing, Eq. (\ref{horizonrho}) demonstrates how an 
experimental realization of a dielectric analogue horizon might be 
understood in terms of the parameters $n,\beta,\omega$ which describe 
the dielectric and kinematic properties of the medium. We expect 
that such horizons can be observed for stationary inhomogeneous 
kinematic configurations of the dielectric matter flow.  The cylindrical 
configuration is favored due to the combined effect of both the 
refraction index $n$ and also of the vorticity $\omega$ of the medium 
which yields a smaller threshold for the radial velocity $\beta$.  
The horizon structures in both cases are shown to be independent 
of the presence of nonlinearities in the permittivity tensor.  

As a final remark, 
it is worthwhile to mention that the present scheme
may provide room also for the formation of sonic horizons 
(dumb holes \cite{Unruh}) 
for the cases in which ultra-sonic velocities are achieved.  

In the present paper, the study of light propagation in a 
nonlinear moving dielectric medium has revealed the existence of a 
surface which displays some of the properties of 
the event horizon of a black hole.
We thus confirm the possibility of creating electromagnetic 
analogues of gravitational black holes in laboratory conditions.  

\begin{acknowledgments}
This work was partially supported by the Brazilian research agencies 
CNPq, 
FAPEMIG, and 
FAPESP. 
\end{acknowledgments}


\end{document}